\documentclass[12pt]{article}
\usepackage{amssymb}
\oddsidemargin = \evensidemargin
\def\bbbr{\mathbb{R}}
\def\bbbc{\mathbb{C}}
\def\bbbone{{\mathchoice {\rm 1\mskip-4mu l} {\rm 1\mskip-4mu l}
{\rm 1\mskip-4.5mu l} {\rm 1\mskip-5mu l}}}

\usepackage{colordvi}

\begin{document}
\title{\bf Weak Quantum Theory: \\ Complementarity and Entanglement \\ in Physics and Beyond}

\author{Harald Atmanspacher \\
Institut f\"ur Grenzgebiete der Psychologie und Psychohygiene\\
Wilhelmstr.~3a, D--79098 Freiburg; \\
Max-Planck-Institut f\"ur extraterrestrische Physik \\
D--85740 Garching
\and 
Hartmann R\"omer \\
Institut f\"ur Physik, Universit\"at Freiburg \\
Hermann-Herder-Str.~3, D--79104 Freiburg 
\and 
Harald Walach \\
Institut f\"ur Umweltmedizin und Krankenhaushygiene, \\
Universit\"atsklinikum Freiburg \\
Hugstetterstr.~55, D--79106 Freiburg  }

\date{}

\maketitle

\bigskip\bigskip


\vspace{2cm}

\begin{abstract}

The concepts of complementarity and entanglement are considered with respect to 
their significance in and beyond physics. A formally generalized, weak version of 
quantum theory, more general than ordinary quantum theory of physical systems,
is outlined and tentatively applied to two examples.  

\end{abstract}

\vfill\eject

\section{Introduction} 

Complementarity and entanglement are notions which have become popular through
the significance they received in quantum theory. Nevertheless they were and
are applied in other fields, even beyond physics, as well. There are cases in which
their purely physical meaning is naturally extended, in other cases they are used
in ways making the connnection to physics hard, inscrutable, or even impossible. 

Applying complementarity and entanglement beyond physics, one is faced with
three logical possibilities.
\begin{itemize}
\item Within a strong reductionist approach, one would understand every kind of
complementarity and entanglement as a manifestation of the quantum theoretical 
significance of those notions in an apparently different context.
\item Assuming that the formal scheme of ordinary quantum theory has realizations beyond 
physics, the application of complementarity and entanglement is possible by direct and
complete analogy to ordinary quantum theory. 
\item A weaker assumption is that a generalized version of the formal scheme of ordinary 
quantum theory, in which particular features of ordinary quantum theory are not contained,
should be used in some non-physical contexts. If concepts like complementarity and
entanglement can still be defined in such a generalized scheme, a 
generalization of those notions beyond physics is achieved.    
\end{itemize}

In this contribution we propose a formal way to define the concepts of
complementarity and entanglement in the spirit of the third option. We do this
in a manner allowing a stepwise relaxation of restricting conditions needed for their 
definition in the context of ordinary quantum physics.
The main purpose of this approach is to sketch possibilities for applying the two notions
in a less restricted, but equally precise way. As will be demonstrated, the intended 
areas of possible application beyond the scope of physics cannot be successfully
addressed in terms of either of the first two options mentioned above. In other words,  
the attempt is to generalize the mathematical and conceptual framework of physical quantum theory 
in such a way that the generalized, weak version of the theory is still mathematically 
formulated, but no longer restricted to physics in its traditional scope.

To be a bit more explicit, complementarity will be extended beyond the concept
of non-commuting properties of a quantum system such as momentum and position as 
elements of a C*-algebra. Entanglement, which is tightly related to
complementarity, will similarly be extended 
beyond the concept of (generally) non-local correlations (not interactions) 
between non-commuting properties of quantum systems. In particular, a formal framework will
be outlined that might facilitate using the concepts of complementarity and
entanglement in situations exceeding the limits of physics as a science of the material world.
For instance, the significance of complementarity and entanglement could be explored in 
philosophical, psychological or psychophysical problem areas, without loosing the desirable 
formal rigor and precision. In this context,
the question is which features of physical quantum theory must be relaxed if one wants to 
apply weak versions of quantum theory to those problem areas.  

The paper is organized in the following manner. In Sec.~2, we give some selected examples for 
complementarity and entanglement for physical, psychological, philosophical, and psychophysical
situations. Sec.~3 provides a compact overview concerning the formal and conceptual framework
of quantum theory. We use an algebraic point of view since the algebraic formulation is best
suited for a clear and transparent discussion of conceptual issues. Moreover it is abstract enough
to open the possibility that it can be applied beyond the physical domain. In Sec.~4, a weak 
version of quantum theory, neutral with respect to its application in a specific scientific field, will be
presented. Conditions will be outlined under which weak quantum theory can be stepwise 
restricted in order to recover the ordinary quantum theory used in physics. Sec.~5 will indicate 
tentative applications, not yet worked out in final detail, of the weak version.
The idea is to look for the concrete significance of general features of weak quantum theory in
selected examples.            

\section{Complementarity and entanglement: \\ some examples}  

Complementarity is a concept made popular by Bohr in his attempts to highlight
crucial features of quantum theory. The textbooks by Meyer-Abich (1965), 
Murdoch (1987), and Pais (1991)
provide a lot of details. A special issue of the journal ``Dialectica'',
edited by Pauli (1948), contains articles on complementarity by leading physicists, 
including Bohr himself (Bohr 1948). 

>From a conceptual point of view, Bohr used
the concept of complementarity to indicate a relationship between apparently
opposing, contradictory notions which can be characterized in terms of a relationship
of polarity. Complementary features typically exclude each other, but at the same time
complement each other mutually to give a complete view of the phenomenon under study.
This is nicely demonstrated by the design Bohr once selected for a medal with which he 
was honored: it shows the text ``contraria sunt complementa'', 
accompanied by the Chinese Yin-Yang symbol.  

The various examples which Bohr discussed
as complementary over the years are of different significance and status. 
To study the corresponding differences in detail, it is 
necessary to look somewhat closer at various ``complementary'' 
pairs of notions. In particular, it is worthwhile   
to explore which ones among them are definitely related to entanglement, which
ones are definitely not related to entanglement, and which ones are (presently)
not understood well enough to draw this distinction clearly.

The best formalized examples of complementary pairs of notions are those referring 
to pairs of non-commutative properties of a system, so-called observables.
Well-known examples are position $Q$ and momentum $P$
(with a generally continuous spectrum) or spins in different directions (with only
two discrete eigenvalues). The fact that, for instance, $P$ and $Q$ do not commute is 
formally expressed as a Heisenberg type commutation relation  
\begin{equation}
[P,Q] = PQ - QP = i \hbar \bbbone ,
\end{equation}
where $\hbar$ is the Planck action $h$ divided by $2\pi$. For more details
see, e.g., Jammer (1974) and references given there.   

The non-commutativity or incompatibility of observables is at the heart of
the non-Boolean structure of quantum theory, and as such it is a major precondition
for situations in which states of systems are entangled. 
Entanglement characterizes the fact that a system in a pure state in general
cannot be simply decomposed into subsystems with pure states. In a certain
sense, such subsystems do not exist a priori but must be generated by appropriate
procedures.     
This has conceptual consequences first pointed out by Einstein et al.~(1935); 
the term entanglement itself was coined by Schr\"odinger (1935). 

The theoretical arguments 
(Bell 1964) and experimental results (Aspect et al. 1982) which, beyond any 
reasonable doubt, confirmed the entangled (holistic) 
characteristics of quantum systems were based on spin-1/2 systems, i.e.
spin measurements on photons. The crucial empirical feature in this context 
are so-called nonlocal (holistic) correlations between two photons. Popular
misconceptions notwithstanding, it is illegitimate to 
interpret these correlations due to causal interactions between the photons.  

In particular situations, the relationship between energy and time can also
be considered as complementary in the sense of non-commutative observables.
Although traditional quantum theory was not general enough to enable
a formal incorporation of corresponding observables, later developments
have shown that energy and time (and related observables) can be rigorously
treated as non-commutative in a more general framework. Major progress in this
respect has been achieved in the theory of stochastic and ergodic systems
(Tj\o stheim 1976, Gustafson and Misra (1976), Misra (1978).

If observables are non-commutative, this implies uncertainty 
relations (such as Heisenberg's uncertainty relations) between them. 
For position and momentum, Heisenberg's uncertainty relation reads:
\begin{equation}
\Delta p \Delta q \ge \hbar/2
\end{equation}
The meaning of this relation is that a quantum system cannot be in a state
in which both $P$ and $Q$ have dispersion-free (definite) expectation values.
The origin of relation (2) is thus of ontic character and goes
beyond epistemic problems of measurement errors or computation errors. 

In ordinary quantum theory, $P$ and $Q$ are maximally incompatible in 
the sense that for every eigenstate of $Q$ all values of $P$ are equally probable and
vice versa. In our approach, we shall call two observables complementary whenever they 
do not commute, even if they are only incompatible rather than maximally incompatible.
In quantum theory, this means that there are pure eigenstates of one of the observables
which are not eigenstates of the other one. It will turn out that this more general notion
of complementarity is suitable for generalizations beyond physics.

The existence of an uncertainty
relation does not necessarily imply that it originates from non-commutative
quantum observables. There are classical observables whose
uncertainty relations are simply due to Fourier reciprocity. 
For instance, engineers have
known for long that the classical bandwidth $\Delta \omega$ (energy) and the 
classical duration $\Delta t$ (time) of a signal satisfy an uncertainty relation 
\begin{equation}
\Delta \omega \Delta t \ge 1/2
\end{equation} 
which is not restricted to ordinary quantum systems 
and does therefore not  imply ordinary quantum entanglement. This situation
will be discussed in Sec.~5.1, see particularly the context of Eq.~(41). 

In addition to non-commutative properties of physical systems,
it is also possible to formalize particular kinds of descriptions
of physical systems in a non-commutative manner. For instance, one can show
that a description of the temporal evolution of a system in terms of a
Liouville operator $L$ and an information theoretical description of the same
system in terms of an information operator $M$ are complementary in the sense that 
\begin{equation}
[L,M] = LM - ML = i K \bbbone
\end{equation}
where $K$ is the (dynamical) Kolmogorov-Sinai entropy of the system (Atmanspacher
and Scheingraber 1987).

Such a complementarity of dynamical descriptions is a special case within
the broader class of deterministic versus statistical descriptions, whose
distinction can be related to that of ontic versus epistemic descriptions
(Scheibe 1973, Primas 1990).  
In contrast to corresponding formalized kinds of complementarity, there are other pairs
of descriptions whose ``complementarity'' is not formally backed up to a comparable
degree. A historical example in quantum physics is the relationship between wave-oriented 
and particle-oriented descriptions. Furthermore, beyond the limits of physics,
there are complementarities of substance and form, of allopoietic and autopoietic
systems, of statistical and deterministic descriptions, of efficient and final 
causation, and many others.
For all of them, it would be as difficult as interesting to show whether 
they are related to some kind of entanglement.

Leaving the natural sciences, things become even more complicated, and
formal approaches are, at least at present, totally lacking. Nevertheless,
e.g., the cognitive sciences and 
psychology offer a multitude of examples which might refer to 
complementarity. The relationship between conscious and unconscious processes
(Jung 1971, Pauli 1954),
between Jung's psychological types (thinking and feeling, intuition and
sensation; Jung 1921), between substantive and transitive mental states
(James 1950), between bi- or multistable states of perception
(Plaum 1992, Kruse and Stadler 1995) and between multiple 
personalities (Jordan 1947) are all candidates for complementary
relations which call for more detailed investigation.
    
Bohr considered the concept of complementarity as so fundamental and
widely applicable that he even used it to characterize philosophical
or philosophy-related problems (cf.~Bernays 1948). 
Three often quoted examples refer to
the definition versus the usage of terms, clarity versus truth,
and goodness versus justice. The complementary pair of 
confirmation and novelty has been proposed for a suitable definition of
meaning in terms of pragmatic information (Weizs\"acker 1974, 
Kornwachs and Lucadou 1985). Probably one of the most far-reaching
applications of complementarity, however, concerns the relation between
mind and matter or, respectively, between mental and material observables
of systems. 

One crucial problem area in this respect is the relationship between 
the psychological experience of mental activity and the (neuro-)physiological
brain processes without which such experience does (most likely) not exist
(cf.~Chalmers 1996).       
This example shows in a particularly clear manner how problematic it is to 
prematurely interpret such relationships in terms of causality. 
The basic concept for a complementary relationship between mind and matter
is that of correlations, i.e., ``neural correlates of consciousness'',
between the considered notions. The question whether there are causal
interactions on top of these correlations is, of course, important. 
However, assigning causality too quickly can lead to ill-posed questions.
It is illustrative to consider many features of the debate between adherents 
of ``mind over matter'' 
versus those of ``matter over mind'' from this point of view.

Another, even more speculative, approach to the relationship between mind
and matter was put forward by Jung and Pauli (1952). Inspired by the old philosophical
concept of psychophysical parallelism, such as in Leibniz's
philosophy, Jung and Pauli explored the idea of a complementary relationship
between mind and matter in a very broad sense. An essential aspect of their 
speculations was a reality behind
(or beyond) those two realms which is, e.g. for epistemological purposes,
split into a mental and a material domain. This split (sometimes called Cartesian
cut) destroys the primordial wholeness of the background reality, and 
``synchronistic'' correlations 
between mind and matter remain as remnants of the lost wholeness. Such a scenario
is obviously inspired by the quantum theoretical conception of entanglement
(see, e.g., Atmanspacher and Primas 1996, Walach and R\"omer 2000, Atmanspacher 2001).
Concrete and detailed indications concerning the substance of such a scenario 
are, at least to our knowledge, not available so far.

In view of all these examples, the question arises whether it is possible to
generalize the standard quantum theoretical framework in such a way that 
complementarity and entanglement might be useful concepts in a broader context.
Moreover, since the two concepts are not identical with each other, it 
is worthwhile to formally explore the conditions which particular situations must  
satisfy to allow us to talk about complementarity and entanglement. 
For this purpose, the next section gives a brief outline of the essential
formal and conceptual features of standard quantum theory from an algebraic perspective.  
Subsequently, it will be studied which of those features can be relaxed
without losing the structures necessary for complementarity and entanglement.

\section{Algebraic quantum theory in a nutshell} 

The algebraic formulation of quantum theory is the most appropriate framework for discussing its 
formal and conceptual structure
and possible generalizations. In this section, we give a brief overview of
algebraic quantum theory in order to provide a solid foundation for the following sections. More
comprehensive and detailed accounts can be found in standard textbooks and monographs, such as 
Haag (1996), Piron (1976), Primas (1983), or Thirring (1981).                             

The fundamental notions for the description of a quantum system $\Sigma$ are those of 
{\it observables} and {\it states}. An observable is any property of the system $\Sigma$ which can -- at least in principle -- be measured in
a reproducible way. To every observable $A$ belongs a set $specA \subset \bbbc$ of
complex numbers, the set of possible results of a measurement of $A$. The observables of $\Sigma$ generate
a $C^*$-algebra $\cal A$ which is called the {\it observable algebra} of $\Sigma$. The system $\Sigma$ can
be in different physical states, and a physical state $z$ determines the probability distributions for the
 measured values of any observable $A$. In quantum theory, states are positive linear functionals on $\cal A$.

Let us explain these notions in more detail. First of all,
the observable $C^*$-algebra $\cal A$ is an algebra over the complex numbers $\bbbc$, which means that
 addition and multiplication of elements in $\cal A$ as well as multiplication with complex
 numbers are defined such that for $A,B,C \in {\cal A}, \alpha, \beta \in \bbbc$:

\begin{itemize}
\item[A1] $A + (B + C) = (A + B) + C$
\item[A2] There is a zero element 0 with $0 + A = A + 0 = 0$.
\item[A3] To every $A$ there is an opposite element $-A$ such that $A + (-A) = A - A = 0$.
\item[A4] $A + B = B+A$
\item[A5] $1 \cdot A = A$
\item[A6] $\alpha(\beta A) =( \alpha\beta )A$
\item[A7] $(\alpha + \beta)A = \alpha A + \beta A$
\item[A8] $\alpha(A + B) = \alpha A + \alpha B$
\item[A9] $A(BC) = (AB)C$
\item[A10] There is a neutral element $\bbbone \in \cal A$ with $\bbbone A = A\bbbone = A$.
\item[A11] $(\alpha A)B = A(\alpha B) = \alpha AB$
\item[A12] $A(B + C) = AB + AC, (B + C)A = BA + CA$
\end{itemize}

\noindent Moreover, $\cal A$ is a star-algebra, i.e., there is an involution $A \mapsto A^*$ with:

\begin{itemize}
\item[S1] $(\alpha A + \beta B)^*  = \overline\alpha A^* + \overline\beta B^*$ ($\overline\alpha$ is the
 complex conjugate of $\alpha \in \bbbc$)
\item[S2] $(AB)^* = B^*A^*$
\item[S3] $(A^*)^* = A$
\end{itemize}

\noindent $\cal A$ is also a {\it Banach star-algebra}, which means that there is a norm function
$A \mapsto || A || \in \bbbr$ with

\begin{itemize}
\item[B1] $||A|| \ge 0, \ ||A|| = 0 \ \Leftrightarrow A = 0, \ ||\alpha A|| = |\alpha| \  ||A||$
\item[B2] $||A + B|| \ \le \  ||A|| + ||B||$
\item[B3] $||AB|| \ \le  \  ||A|| \  ||B||$
\item[B4] $||A^*|| \ = \  ||A||$
\item[B5] $\cal A$ is complete with respect to the norm $|| \cdot ||$, i.e., every sequence $A_n$ with
 $||A_m - A_n || < \varepsilon$ for $m,n \ge N(\varepsilon)$ converges to a (unique) element of $\cal A$.
\end{itemize}

\noindent Finally, the so-called {\it $C^*$-condition} for $\cal A$ means:

\begin{itemize}
\item[C1] $||A^*A|| = ||A||^2 \quad$ (B4 follows from C1) 
\end{itemize}

\noindent A state $z$ is a (continuous real) functional $A \mapsto z(A) \in \bbbc$ on $\cal A$
with:

\begin{itemize}
\item[Z1] $z(\alpha A + \beta B) = \alpha z(A) + \beta z(B), \quad z(A^*) = \overline{z(A)}$ 
\item[Z2] $z(A^*A) \ge 0$ 
\end{itemize}

The reader will notice that we also admit an ``impossible'' zero state $z = 0$. The set of
 all states is {\it convex}, hence with $z_1$ and $z_2$ also

\begin{equation}z = \alpha z_1 + (1 - \alpha)z_2, \quad 0 \le \alpha \le 1
\end{equation}
is a state.
A state $z$ is called {\it pure}, if it does not admit a non-trivial decomposition of type (5). Pure
states contain maximal information about the system $\Sigma$.

For every state $z \neq 0$ we can define {\it expectation values} of observables $A$:
\begin{equation}
E_z (A) = {z(A) \over z(\bbbone)}.
\end{equation}
$ E_z (A)$ is the mean of measured values of $A$ in the state $z$, if the (reproducible) measurement
 of $A$ is repeated many times in the same state $z$.

In quantum theory, $specA$, the set of possible measured values of $A$ is given by those $\alpha \in \bbbc$,
 for which $(A - \alpha {\bbbone})$ has no inverse in $\cal A$. Moreover, only self-adjoint
  elements of $\cal A$ are normally admitted as observables:
\begin{equation}
A^* = A;
\end{equation}

The {\it uncertainty} $\sigma_A^z$ of a self-adjoint observable $A \in \cal A$ in a state $z \neq 0$ is defined as
\begin{equation}
(\sigma_A^z)^2 = E_z((A - E_z(A))^2) = E_z(A^2) - E_z(A)^2 \ge 0.
\end{equation}
One can derive a general uncertainty relation
\begin{equation}
\sigma^z_A \sigma^z_B \ge {1 \over 2} \ | \ E_z (AB - BA) \ |.
\end{equation}

In quantum theory, as opposed to classical theory, the observable algebra is not commutative. This means
that, in general, $AB - BA \ne 0$, and there is no state in which all uncertainties vanish.
Commuting observables with $AB = BA$ are called {\it compatible}, non-commuting observables with
$AB \neq BA$ are called {\it incompatible} or {\it complementary}.

{\it Propositions} are special observables, whose measured values can only
be ``yes'' $\widehat{=} \ 1$ or ``no'' $\widehat{=} \ 0$. They are given by elements $P \in \cal A$ with 
\begin{equation}
P^* = P,  \  P^2 = P.
\end{equation}
For $z \neq 0$, $E_z(P)$ is the probability that $P$ is measured as ``yes'' in the state $z$.

To every proposition $P$ we can associate a {\it negation} ${\overline P} = \bbbone - P$ with
\begin{equation}
E_z({\overline P}) = 1 - E_z (P).
\end{equation}
\noindent The {\it conjunction} of two propositions $P_1$ and $P_2$ is given by the proposition
\begin{equation}
P_1 \wedge P_2 = \lim_{n\rightarrow\infty} (P_1P_2)^n,
\end{equation}
and the {\it adjunction} is defined as
\begin{equation}
P_1 \vee P_2 = \bbbone - (\bbbone - P_1)  \wedge  (\bbbone - P_2) = \overline{{\overline P_1}
\wedge  {\overline P_2}}.
\end{equation}
\noindent We find
\begin{eqnarray}
P_{1,2} (P_1 \wedge P_2) = (P_1 \wedge P_2) P_{1,2} = P_1 \wedge P_2 \nonumber, \\
P_{1,2} (P_1  \vee P_2) = (P_1 \vee  P_2)P_{1,2} =  {P_{1,2}}.
\end{eqnarray}
\noindent For compatible $P_1, P_2$ we simply have

\begin{eqnarray}
P_1 \wedge P_2 = P_1 P_2 \nonumber \\
P_1 \vee P_2 = P_1 + P_2 - P_1 P_2
\end{eqnarray}

The {\it spectral theorem} states that every self-adjoint observable $A$ can be equivalently
represented as the adjunction of propositions which are mutually compatible and compatible with $A$.

In the usual formulation of quantum theory, observables are operators and states are density 
matrices on a
{\it Hilbert space} $\cal H$. In the (more general) algebraic formulation, such Hilbert space formulations can be
 recovered by the so-called {\it GNS-construction}.

Every state $z \neq 0$ defines a vanishing ideal
\begin{equation}
I_z = \{C \in {\cal A} \ | \  z(C^*C) = 0\}.
\end{equation}
The quotient algebra ${\cal A}  / I_z$, together with the scalar product
\begin{equation}
<[A], [B]>_z = z(A^*B)
\end{equation}
on the equivalence classes $[A] = A + I_z, [B] = B + I_z$, then gives rise to a Hilbert space ${\cal H}_z$
 and to a representation
\begin{equation}
A[B] = [AB]
\end{equation}
of $\cal A$ on ${\cal H}_z$.

For what follows it is important that every observable $A \in \cal A$ acts on the set $Z$ of all states:
 to every state $z \in Z$ and every observable $A \in \cal A$ we can associate a state
\begin{equation}
\rho_A (z)
\end{equation}
defined by
\begin{equation}
(\rho_A(z)) (B) = z (A^*BA).
\end{equation}
\noindent Obviously,
\begin{equation}
\rho_{A_1A_2}(z) = \rho_{A_1} \rho_{A_2} (z).
\end{equation}

We can identify $A$ and $\rho_A$ up to a complex phase, which is 
generically fixed, if $specA$ is known.
The action of observables on states corresponds to the active interpretation of observables
as operations changing the state of a system. For a proposition $P$ and a state $z$ with $z(P) \neq 0, 
\rho_P(z) =: Pz$ is a state in which $P$ is true with certainty:

\begin{equation}
E_{Pz}(P) = {Pz(P) \over Pz(\bbbone)} = {z(P^3) \over z(P^2)} = 1.
\end{equation}

Application of $P$ corresponds to verification of $P$. One remarkable feature of quantum theory is its
{\it holistic} character. If a quantum system $\Sigma$ is composed of two subsystems $\Sigma_1$ and
 $\Sigma_2$, then the state of $\Sigma$ is in general not determined by the states of $\Sigma_1$ and
  $\Sigma_2$. The reason for this resides in the non-commutativity of the algebra $\cal A$ of $\Sigma$.
  $\cal A$ contains subalgebras ${\cal A}_1$ and ${\cal A}_2$ refering to the subsystems $\Sigma_1$ and
   $\Sigma_2$. If $\Sigma_1$ and $\Sigma_2$ are well-separated, ${\cal A}_1$ will commute with ${\cal A}_2$.
    Nevertheless, there will be observables $A \in \cal A$ of the total system which are
    incompatible with observables in ${\cal A}_1$ and/or ${\cal A}_2$. Intimately related to this,
    there exist {\it entangled states} of the total system, in which the values of observables in
    ${\cal A}_1$ and ${\cal A}_2$ are undetermined but correlated without any interaction between
     $\Sigma_1$ and $\Sigma_2$. These correlations cannot be used for transmitting information between
     $\Sigma_1$ and $\Sigma_2$. If, in addition, the algebras ${\cal A}_1$ and
     ${\cal A}_2$ are non-commutative, {\it Bell's inequalities} for correlations between measured
      values of observables for $\Sigma_1$ and $\Sigma_2$, which must be satisfied under the 
hypothesis of local realism, can be shown to be violated. This is the case in quantum theory.

Under the assumption that there is no instant interaction-at-a-distance between $\Sigma_1$ and
$\Sigma_2$, the (experimentally observed) violation  of Bell's inequalities means that a realistic
interpretation, to the effect that the outcome of measurements on $\Sigma_1$ and $\Sigma_2$ is predetermined
 by objective features such as local hidden parameters already before measurement, is excluded. The indeterminacy
 of quantum theory is not {\it epistemic}, i.e. due to incomplete knowledge or inevitable perturbations of
 the state of the quantum system $\Sigma$, but {\it ontic}. This fact can and should be interpreted as a
 consequence of the holistic character of quantum theory.

Separation and composition are problematic operations in complex systems. Quite a basic
example is the separation
given by the {\it epistemic splitting} of a physical system into an observing and an
 observed subsystem. A quantum theoretical analysis of the measuring process reveals that the stochastic
 character of quantum theory can be attributed to a neglect of holistic correlations between observing
  and observed subsystem if the system is in an entangled state.
Entangled states of complex systems have a tendency to evolve
 into decoherent states which are effectively indistinguishable from an incoherent superposition of separable
  states.

\section{Weak quantum theory}

In this section, we outline some assumptions under which particular features of quantum theory can
be generalized to a framework broader than that of ordinary quantum theory. For such a framework,
we expect the notions of systems, observables and states to remain
valid and meaningful. 

A system $\Sigma$ is considered as a part of reality in a very general sense, i.e. it can be the 
object of attention and investigation beyond the realm of ordinary quantum theory, possibly even beyond 
the limitations set by the concept of a material reality. Even though the isolation of parts of reality 
is expected to be a problematic operation,
its possibility, at least in some approximate sense, is the prerequisite for any act of cognition and, in
fact, already implicit in the epistemic split between subjects and objects of cognition.

An observable is any (more or less) meaningful property of the system $\Sigma$ which can be investigated
in a given context. Non-trivial observables must exist, whenever $\Sigma$ has enough internal structure
to be a possible object of a meaningful study. To every observable $A$ there should belong a set
 $specA$ of possible results of an investigation of $A$. In the general case addressed in this section,
the relation between $A$ and $specA$
 will be different from the quantum theoretical situation described in the previous section.

It must at least be conceivable that the system $\Sigma$ exists in different states. Different
states should reflect themselves in different outcomes of observations associated to observables $A$.
 Even if the system $\Sigma$ de facto always is in the same state $z$, it must be possible to
 conceive it in other states. (Otherwise, nothing could be learned about $\Sigma$.) The
  possibility of different states is indispensible for discussing stability criteria for the system 
$\Sigma$,
   which has to maintain its identity under ``unsubstantial'' changes.

In addition, the notion of a state also has epistemic aspects, reflecting various degrees of knowledge
about $\Sigma$. As in ordinary quantum theory, we call a state pure if it contains maximal information 
about $\Sigma$. Normally, the state of the system $\Sigma$ will change in the course of time, under the
 influence of other systems and as a consequence of being observed and investigated.

As in ordinary quantum theory, we associate a set $\cal A$ of observables and a set $Z$ of states to 
every system
$\Sigma$. Here $\cal A$ and $Z$ are meant to be sets in a naive sense, not necessary in the sense of
axiomatic set theory. Our task will be to investigate the general structures of $\cal A$ and $Z$.

The first property of $\cal A$ we want to formulate is \\

\smallskip
{\it Axiom I}: To every observable $A \in \cal A$ belongs a set $specA$, the set of possible outcomes of
a ``measurement'' of $A$. \\
\smallskip

In the previous section we saw that quantum observables can be identified with functions
 $A : Z \rightarrow Z$ on the set of states.
This fact, which underlines the active, operational character of observations, should be valid more
generally. We thus formulate \\

\smallskip
{\it Axiom II}: Observables are (identifyable with) {\it mappings} $A: Z \rightarrow Z$, which associate
to every state $z$ another state $A(z)$. \\
\smallskip

Axiom II implies that observables can be composed as maps on $Z$, where the map $AB$ is defined by first
applying $B$ and then $A$. We shall assume \\

\smallskip
{\it Axiom III}: With $A$ and $B$, also $AB$ is an observable. \\

\smallskip
A direct consequence of Axiom III is the {\it associativity} of the composition of observables:

\begin{equation}
A(BC) = (AB)C
\end{equation}
\noindent Moreover, we can postulate  \\

\smallskip
{\it Axiom IV}: There is a unit observable $\bbbone$ such that $\bbbone A = A \bbbone = A \  \forall  \ A \in \cal A$.\\
\smallskip

$\bbbone$ is the operation on $Z$ which does not change any state, it corresponds to a proposition which
is always true, so $spec\bbbone = \{{\rm true}\}$.
Axioms II--IV mean that the set of observables has the structure of a {\it monoid}, which is also
called a {\it semigroup with unity} or an {\it associative magma with unity}.

For formal completeness we also need an ``impossible'' {\it zero state} $z = o$ and a {\it zero observable} 0 with
$spec0 = \{{\rm false}\}$, which corresponds to an always false proposition.\\

\smallskip
{\it Axiom V:} There are a zero state $o$ and a zero observable 0 such that 
\begin{eqnarray}
0(z) = o \ \forall\  z \in Z ,  \nonumber \\
A(o) = o \ \forall\  A \in \cal A, \\ 
A0 = 0A = 0 \ \forall\  A \in \cal A. \nonumber 
\end{eqnarray}

One may wonder about the {\it addition} of observables which, after all, plays an important role in
ordinary quantum theory. But even there, the operational definition of $A + B$ is problematic;
for instance there is no general strategy to construct a measuring device for $A + B$. Nevertheless it 
should be mentioned that Jordan algebras, an early attempt to generalize ordinary quantum theory
from non-commutative, associative algebras to commutative, non-associative algebras, are explicitly
constructed on the basis of an addition of observables (for more details see Primas 1983).         

In the generalized framework addressed here, there is no evident place for the addition of observables. 
As a consequence, the set of states in weak quantum theory cannot be presupposed to be convex as in ordinary
quantum theory. This relates to the fact that a probability interpretation is not feasible within 
the general framework of weak quantum theory (but can be implemented by restricting its generality,
see below).

There is no reason to assume commutativity, $AB = BA$, for all $A$, $B \in \cal A$. Rather there will
 be both commutative (compatible) and non-commutative (incompatible) pairs of observables, 
depending on whether $AB = BA$ or $AB \neq BA$.
 This means that the monoid structure of $\cal A$, however poor and general, contains 
complementarity and entanglement as essential features of quantum theory.

A simple model obeying axioms II--IV can be obtained by 
identifying $Z$ with the states of a rigid body in space and $\cal A$ with the
set of motions. This indicates that additional structure is required for
 a reasonable generalization of ordinary quantum theory. 
Such additional structure is, indeed, at hand,
 because there are {\it propositions} $P$ among the observables $\cal A$  which play a distinguished role.
 A proposition $P \neq 0, \bbbone $ is an observable whose outcome is either true or false:
\begin{equation}
specP = \{{\rm true, false}\}\  {\rm for} \  P \neq 0, \bbbone
\end{equation}

\noindent Moreover, to every proposition $P$ there must be a negation $\overline P$, which gives
``false'' if and only if $P$ gives ``true''.

We now give a few rather evident axioms assumed to hold for propositions. \\
\smallskip
{\it Axiom VIa}:
\begin{eqnarray}
P^2 = P, \nonumber \\
\overline{\overline P} = P, \  \overline{\bbbone} = 0, \\
P\overline P = \overline P P = 0 \nonumber .
\end{eqnarray}
\smallskip

For {\it compatible} propositions $P_1, P_2, P_1 P_2 = P_2 P_1$ we can define a 
{\it conjunction}
\begin{equation}
P_1 \wedge P_2 = P_2 \wedge P_1 =P_1P_2
\end{equation}
and an {\it adjunction}
\begin{equation}
P_1 \vee P_2 = \overline{\overline P_1 \overline P_2} = P_2 \vee P_1
\end{equation}
with the properties
\begin{eqnarray}
P\wedge P = P\vee P = P, \nonumber \\
P_1 \wedge (P_2 \wedge P_3) = (P_1 \wedge P_2) \wedge P_3,  \nonumber \\
P_1 \vee (P_2 \vee P_3) = (P_1 \vee P_2) \vee P_3, \nonumber \\
P_1 \wedge (P_1 \wedge P_2) = (P_1 \wedge P_2) \wedge P_1 = P_1\wedge P_2,  \nonumber \\
P_1 \vee (P_1 \vee P_2) = (P_1 \vee P_2) \vee P_1 = P_1\vee P_2,  \\
0 \wedge P_1 = P_1 \wedge 0 = 0, \nonumber \\
\bbbone \wedge P_1 = P_1 \wedge \bbbone = P_1, \nonumber \\
0 \vee P_1 = P_1 \vee 0 = P_1, \nonumber \\
\bbbone \vee P_1 = P_1 \vee \bbbone = \bbbone. \nonumber
\end{eqnarray}

Moreover, we postulate the meaning of $P$ as verification. \\
\smallskip
{\it Axiom VIb:} If $P(z) \neq o$, then $P(z)$ is a state in which $P$ is true with certainty.
\smallskip

Finally, we formulate an axiom replacing the spectral theorem of ordinary quantum theory. Every
observable $A$ should be equivalent to a set of mutually exclusive  propositions. More precisely,
 let $A$ be an observable and $\alpha \in specA$. $A_\alpha$ denotes the proposition that the outcome
  of a measurement of $A$ is $\alpha \in specA$. Then we have \\
\smallskip
{\it Axiom VIc:}
\begin{equation}
A_\alpha A_\beta = A_\beta A_\alpha = 0 \ {\rm for} \  \alpha \neq \beta, \quad A A_\alpha = A_\alpha A, \quad
\bigvee_{\alpha \in specA} A_\alpha = \bbbone.
\end{equation}
$A$ and $B$ are compatible if and only if $A_\alpha$ and $B_\beta$ are compatible for all
$\alpha \in specA$ and $\beta \in specB$.
\smallskip

In general, incompatible observables do not have simultaneous definite values.

Although the generalized {\it weak quantum theory} as defined by axioms I--VI is considerably
 weaker than ordinary quantum theory, they share the following two characteristic features.

\begin{itemize}
\item Incompatibility and complementarity arise due to the non-commutativity of the multiplication
 of observables.
\item Holistic correlations and entanglement arise if for a composite system observables pertaining
to the whole system are incompatible with observables of its parts. 
\end{itemize}

In the latter context, it should be emphasized that weak quantum theory itself
refers to the description of the system as a whole. Any identification of parts or subsystems
implies a specific choice of representation in terms of partial monoids. This choice remains
open in the general framework of weak quantum theory. In weak quantum theory, the absence
of  a vector space structure implies that there is no tensor product construction for the set of
observables of a composite system. In general, we can only expect: 
\begin{eqnarray}
{\cal A} \supset {\cal A}_1 \times {\cal A}_2, \  Z \supset Z_1 \times Z_2,  \\
{\cal A}_1 (Z_1)  \subset Z_1, \  {\cal A}_2 (Z_2)  \subset Z_2.
\end{eqnarray} 

A similar remark applies to the specific form of the dynamical evolution of (sub-)systems
in weak quantum theory. The dynamics of a system is generally described
by a one-parameter (semi-)group of endomorphisms. The process generating subsystems 
(e.g., by measurement) and the dynamics of interacting subsystems depends on 
details of the considered system and its decomposition.
  
There are other features of weak quantum theory which are not shared by ordinary quantum theory.

\begin{itemize}
\item There is no quantity like Planck' s constant $h$ which in ordinary quantum
theory quantifies the degree of non-commutativity of two given observables. This indicates 
that in the generalized, weak theory, complementarity
and entanglement are not restricted to a particular degree of non-commutativity as it is 
the case for ordinary quantum mechanics.
\item Since the addition of observables is not defined in the general framework of weak quantum 
theory, there is no convex set of states, there are no linear expectation value functionals, and there is 
no probability interpretation. 
Probability distributions on the sets $specA$ do not occur and are not calculable in
weak quantum theory. As a matter of fact, the mere concept of probability will be absent in many
situations (e.g., in an exploration of a work of fine art or of the intensity of an emotion).
\item There is no way to generalize Bell's inequalities up to the general framework of 
weak quantum theory, and there is  no way to argue
that complementarity and indeterminacy in weak quantum theory are of ontic rather than epistemic
nature. On the contrary, one would expect them to be of rather innocent epistemic origin in 
many cases,
 for instance, due to incomplete knowledge of the system or uncontrollable perturbations by observation.
\end{itemize}

Axioms I--VI, characterizing weak quantum theory, can be regarded as minimal requirements
for a meaningful general theory of observables and states of systems. Between the weak version of
quantum theory and its ordinary version, there are intermediate theories which can be obtained by 
enriching the axioms stepwise. Let us first discuss enrichments of the propositional axiom VI. Subsequently
we shall add a probability interpretation of states.

One evident option is to postulate that the conjunction and adjunction of propositions is
also defined in the less intuitive case of {\it incompatible} $P_1$ and $P_2$ such that
propositions $P_1 \wedge P_2$ and
$P_1 \vee P_2 = \overline{{\overline P_1} \wedge {\overline P_2}}$ always fulfil the conditions 
of equations (29). In addition, it is natural to postulate
\begin{eqnarray}
P_1 \wedge (P_1 \vee P_2) &=& (P_1 \vee P_2) \wedge P_1 = P_1,     \nonumber \\
P_1 \vee (P_1 \wedge P_2) &=& (P_1 \wedge P_2) \vee P_1 = P_1 \wedge P_2.
\end{eqnarray}
The stronger distributivity condition
\begin{eqnarray}
P_1 \wedge (P_2 \vee P_3) &=& (P_1 \wedge P_2) \vee (P_1 \wedge P_3),     \nonumber \\
P_1 \vee (P_2 \wedge P_3) &=& (P_1 \vee P_2) \wedge (P_1 \vee P_3),
\end{eqnarray}
is not even satisfied in ordinary quantum theory. If every propositional subsystem generated 
by two compatible propositions with $P_1\wedge P_2 = P_1$ and their negations is Boolean, 
then (modulo some technical complications) the propositional system is already isomorphic 
to a system of orthogonal projectors in a Hilbert space (Piron 1976, Thirring 1981).
This Boolean property does not follow from axioms I -- VI.

It cannot be prescribed in general which
of these additional assumptions are applicable in a concrete situation. It will be unavoidable to 
consider details of the given context for corresponding decisions. 

For a probability interpretation  of states, one does not loose much by assuming $specA \in \bbbc$,
 because it is very plausible that the set of outcomes of $A$ can be mapped into the complex numbers
 in  a one-to-one way. Assuming this, the existence of a probability interpretation amounts to
 postulating for every $z \neq 0$ the existence of an expectation value functional
\begin{eqnarray}
E_z: {\cal A} \rightarrow \bbbc, \nonumber \\
A \mapsto E_z(A) \in \bbbc, 
\end{eqnarray}
with
\begin{equation}
E_z(\bbbone) = 1.
\end{equation}

The existence of an expectation value functional has far reaching consequences
\begin{itemize}
\item Addition of observables and multiplication of observables with complex numbers can now be
defined by postulating
\begin{equation}
E_z(\alpha A + \beta B) = \alpha  E_z(A) + \beta E_z(B)
\end{equation} 
for all $E_z$.
Axioms A1--A10 are fulfilled, A11 seems to be natural, less so A12.
\item Being the mean value of a probability distribution, $E_z(A)$ has to obey reality and positivity
 conditions. The only evident way to achieve this is the introduction of a star-involution
 $A \rightarrow A^*$ with the properties S1--S3. (S2 has to hold, because $A^*A$ has to be
 self-adjoint also if $A$ and $A^*$ do not commute.) Reality and positivity of $E_z$ mean
  that Z1 and Z2 have to hold for all $E_z$.

\item The set of all expectation value functionals will be convex. Pure states can be defined as in
ordinary quantum theory.
\end{itemize}

The axioms B1--B5 are almost mandatory if one assumes that $\cal A$ can be topologized by a norm.
 The $C^*$-axiom C1 is least intuitive. Assuming C1 on top of the other axioms A, S, B, Z, ordinary
 quantum theory can be recovered. As mentioned above, only a detailed analysis of the concrete 
situation can decide which axioms are fulfilled.

\section{Complementarity and entanglement in weak \\ quantum theory: two applications}

In this section, we outline two examples for the application of weak
quantum theory. As mentioned in the preceding section, it should be possible 
to construct frameworks less restrictive than ordinary quantum theory 
but more restrictive than the weak version. Our first example, the complementarity
of different types of dynamical descriptions of physical systems, addresses precisely 
such a situation. This example is particularly interesting since it can be presented in
a fairly well formalized manner.   

The second example, addressing transference and countertransference
phenomena in psychology, will be discussed in an entirely qualitative and informal
way. The basic complementarity in this example is that of conscious and unconscious 
processes, hence the relevant states and observables are mental, not material. 
It is likely that this example refers to the minimal set of
axioms given in the preceding section.     

There are additional possibilities to illustrate the applicability of weak quantum theory.
An example which is intended to be worked out in detail elsewhere addresses 
a complementarity between the effects of placebo substances 
and ``true'' treatment substances in double-blind clinical trials. 
It is understood to refer to the effects of such
substances in a purely pharmacological (physiological) sense, i.e., without considering 
possible psychological aspects.

\subsection{Information Dynamics}

Generalizing earlier work by Misra (1978) and Misra et al.~(1979), 
an information theoretical description
of chaotic systems (including K-systems) was found to provide a commutation relation
between the Liouville operator $L$ for such systems and a suitably defined 
information operator $M$ (Atmanspacher and Scheingraber 1987). 
The definition of $L$ is, as usually, given by
\begin{equation}
L\ \rho = i{\partial\over \partial t}\ \rho
\end{equation}
where $L$ acts on distributions $\rho$ which represent the states of a system
in a usual probability space (not in a Hilbert space). The continuous spectrum of 
$M$ derives from the
time-dependent information $I(t)$ which can be gained by measuring 
particular properties of a system at time $t$ in comparison with its
predicted properties:  
\begin{equation} M\ \rho = I(t)\ \rho = (I(0) + Kt)\ \rho
\end{equation}
$K$ is the Kolmogorov-Sinai entropy, a statistical dynamical invariant of the 
system. It is experimentally
available by Grassberger-Procaccia type algorithms (Grassberger and Procaccia 1983).
$K> 0$ only for chaotic systems with intrinsically unstable dynamics.
In an information theoretical interpretation (Shaw 1981), $K$ characterizes the
rate at which the system generates information along its unstable manifolds. 
$Kt$ is the information generated by the system between $t$ and $t=0$.
This means that the accuracy of a prediction decreases with increasing prediction
time. 

In simple cases, the commutator of $L$ and $M$ is just given by
the rate of information generation, namely the Kolmogorov-Sinai entropy:
\begin{equation} i[L,M] = K {\bbbone}
\end{equation}
The two operators commute precisely if the considered system does not
generate information, i.e., if it is intrinsically stable.  
If $K>0$, the dynamical descriptions 
due to $L$ and $M$ are different with respect to the prediction of a future state
of the system. This is a consequence of the increasing uncertainty
in predicting the state of a system as time proceeds. Whenever $K>0$, the state 
$\rho(t)$ of a system cannot be predicted as accurate as initial conditions
have been measured or otherwise fixed at $t=0$.    

The commutation relation of $L$ and $M$ resembles corresponding
commutation relations in ordinary quantum theory, but there are differences.
First of all, since $K$ is explicitly system- and parameter-dependent 
      (i.e.~highly contextual), the ``degree" of non-commutativity   
      of $L$ and $M$ is not universally the same. 
      This situation is at variance with conventional 
      quantum mechanics with $h$ as a universal commutator.  
Moreover, $K$ is a statistical quantity specifying the average flow of information
in chaotic systems, while $h$ is a non-statistical constant of nature. 

As a consequence of relation (40), $L$ and $M$ provide complementary modes of description.
There are two basic features of this complementarity. (i) While $L$
refers to an ontic, completely deterministic description, $M$ refers to an
epistemic, coarse grained description of explicitly statistical nature 
(cf.~Atmanspacher 2000). (ii) While a description
in terms of $L$ is time-reversal symmetric (reversible), this symmetry is broken by a
description in terms of $M$, thus leading to irreversibility.

There is an interesting relation between (40) and another commutation relation
between $L$ and a time operator $T$ introduced by Misra (1978) and Misra et al.~(1979):
\begin{equation} i[L,T] = {\bbbone}
\end{equation}
$T$ is well-defined if $K>0$. Since $L$, in addition to its role as an
evolution operator as in (38), can also be interpreted as an energy difference due to 
$L\rho = [H,\rho]$, (41) indicates a complementarity between energy and time
for chaotic systems. This suggests the idea of a temporal entanglement for 
such systems. This entanglement can be interpreted as a temporal nonlocality
(Misra and Prigogine 1983) due to a coarse grained phase space; for a more detailed
discussion see Atmanspacher (1997). It should be emphasized that this
nonlocality is epistemic and must not be mixed up with the ontic nonlocality of 
ordinary quantum theory.

An interpretation of the commutation relation between $L$ and $M$
in terms of propositions leads to a lattice theoretical analysis. Analogous
to the work of Birkhoff and von Neumann (1936) which pioneered the
non-Boolean logic of quantum theory, such an analysis provides 
basic logical features of information processing systems.
Following an idea by Krueger (1984), it was shown 
that the temporal evolution of information processing systems is
governed by a non-Boolean logic (Atmanspacher 1991a). 
More precisely, the propositional lattice characterizing
such a logic is complemented but not distributive, reflecting the
complementarity of propositions from the perspective of a lattice
theoretical formulation. The non-distributivity
of this lattice, however, shows a subtle but important difference
as compared with the non-distributivity due to ordinary quantum theory.

A fundamental feature of lattices as mathematical structures is the duality
of their properties. Formally this means that each true proposition is
transformed into another true propositon by exchanging the dual operations
defined in lattice theory. It turns out that the subtle difference between the
standard quantum theoretical non-distributivity and the non-distributivity
due to information processing systems precisely accounts for this duality.
While standard quantum theory provides non-distributivity relations of the form
\begin{eqnarray}
&a>(a\wedge b)\vee (a\wedge b')  \nonumber \\
    \wedge\ \  &b>(b\wedge a)\vee (b\wedge a') 
\end{eqnarray}
($a'$ ist the complement of proposition $a$, $b'$ is the complement of 
proposition $b$), 
information processing systems satisfy non-distributivity relations of the form:
\begin{eqnarray}
&a<(a\vee b)\wedge (a\vee b')  \nonumber \\
     \vee\ \   &b<(b\vee a)\wedge (b\vee a') 
\end{eqnarray}
By contrast to (42),  (43) requires only one of the two inequalities 
to be satisfied. A detailed analysis (Atmanspacher 1991a) shows that this 
is indeed crucial for the non-distributivity of information processing systems.   
It is therefore tempting to consider the logics of standard quantum systems and 
of information processing systems
as dual aspects of one underlying non-distributive 
lattice (Atmanspacher 1991b).

\subsection{Countertransference Phenomena}

Freud (1992)  was the first to observe that in the context of a therapeutic relationship strange 
interpersonal experiences can happen which he called transference and 
countertransference. 
Transference normally refers to the fact that the patient activates conflictual relationship 
themes from his past and enacts them in the context of the therapeutic relationship, 
transferring these past experiences into the presence and acting as if the therapist was his 
mother, father, brother or whoever he had the conflictual relationship with. Modern 
therapeutic theories postulate that a potentially helpful therapy will in fact activate 
such past 
experiences in the presence. By countertransference Freud originally meant that 
also within the therapist some potentially conflictual material can be activated by the 
patient, if the therapist is prone to the same problematic pattern as the patient. 

In addition to the traditional and straigthforward meaning of transference and 
countertransference phenomena there is also a more subtle meaning which we will 
discuss in the following. It refers to a therapist's experience of inner states like 
emotions, 
ideas, thoughts, inner images, impulses, needs, phantasies, wishes, which are in fact 
``transferred'' from the patient and reflect the inner state of the patient rather 
than that of the therapist. In clinical practice this countertransference phenomenon is 
used both 
diagnostically and interventionally. Diagnostically it can be a source of direct and intuitive 
information about the inner world of the patient. Thus, if a therapist experiences, in an 
otherwise calm atmosphere with a patient talking serenely about something pleasant which 
he or she has experienced, sudden throngs of aggression or wild phantasies of 
sexual abuse, then he might tentatively isolate these inner experiences as possibly 
belonging to the patient rather than to himself. He might then operate with the 
hypothesis that the calm story of the patient is just the surface, while underneath there 
might lie some material of a more dire nature. 

Depending on the school of therapy the therapist belongs to, the nature of the problem, 
the state of the therapy, and the personality of the patient the therapist might choose 
to express this phantasy explicitly and offer it to the patient as an interpretative 
framework. Alternatively, he might feed back his own inner world, without any 
qualifications, as is often done in Gestalt therapy. Or he may keep it as an information 
which could direct later interventions. In any case, the hypothesis on which therapists 
often base their interventions is that the material they experience themselves 
under certain circumstances derives from the patient rather than from themselves. 
Although it is, strictly speaking, not possible to give clearcut rules as to when material 
is transferred or not, there are some practical rules of thumb. If material feels 
alien or strange to the therapist's state of mind, if it does not ``fit'' with the rest 
of the situation, then there is a good chance that the material is from the patient and 
not from the therapist himself. Therapeutic training within the depth psychological 
schools of therapy places a lot of emphasis and takes great pains and care to sharpen 
the inner awareness for the subtle changes which signal transference processes.

A comparable phenomenon is known from system-therapeutic settings, where family 
constellations are enacted in a group. In some schools the person working on their 
family will, with the help of the therapist, call other group members onto a stage or 
set place. These group members, the so called protagonists, are to take the place of 
the family members. Even family members who are long deceased or whom the patient 
does not know much about might be included in the family picture. It has been 
repeatedly observed that protagonists suddenly experience mental states which belong 
to the person exemplified, without the protagonist explicitly knowing about any 
corresponding details. For instance a protagonist representing a relative who has 
committed suicide, a fact unknown to all persons, might suddenly feel the impulse to 
leave the room. Or another protagonist, who represents someone who had a severe 
war injury might suddenly complain of strong pain, although nobody in the room, 
except perhaps the client, is aware of this fact. System theoretic therapists  
call this phenomenon ``participatory" or ``deputy" perception (Varga v.~Kibed 1998). 

There are reasons to assume that this is a phenomenon of the same type as in a 
classical countertransference 
situation. In both cases it is supposed that someone experiences mental states 
which do not pertain to himself but to someone else, e.g. the patient or some other 
person who is represented in the family setting.
The phenomenon is well known as such, but for a lack of theoretical explanation has not 
found much interest, except for some practical purposes. We propose here to see this 
phenomenon as an example of entangled mental states, a very general situation
addressed by weak quantum theory. 

A crucial condition for successful therapeutic relationships is  
mutual openness. Ideally, both the therapist and the client communicate without 
withholding important or possibly important information. While in the psychoanalytic 
tradition this is a significant part of an explicit agreement, it certainly is also implictly 
true for most therapeutic alliances. But this openness is opposed by all those parts of 
inner material which are not available for conscious processing, either because they are 
subconscious and not known, or because they are not identified as having a particular 
flavor or relevance. 

In order to apply the concepts of weak quantum theory to such situations, we consider 
the entire group of involved people as the system as a whole. The subsystems are the 
individual members of the group with particular emphasis on their  mental (psychological)
variables. The local preparation of ``conscious awareness'' can then
be considered as complementary to a global preparation of material 
which is principally not available, because it is unconscious or irrelevant. 
While the latter corresponds to a global observable of the system as a whole (maybe
referring to some kind of collective unconscious material),
the former corresponds to local observables of subsystems, i.e.~conscious contents 
of the mental system of individuals.

Material which is unconscious cannot be consciously known or even openly communicated, 
and contents which are consciously known or can be communicated openly cannot be unconscious.
In this sense, the concepts of consciousness and unconscious are complementary. They are 
not only opposed to each other, but preclude each other and at the same time are both necessary 
for a complete picture of the overall mental system. 

The process leading from unconscious material to (partly) conscious manifestations of that 
material may be 
conceived as a psychological analogue of the physical process of observation (cf.~Jung 1971). 
In both types of processes, a global state is decomposed into a local state plus an environment, 
where the environment is assumed to include the measuring apparatus. In the psychological
case, this means that a part of consciousness is the analogue of a ``measuring tool'' 
and another, emerging part of consciousness is the analogue of the physical subsystem emerging 
from the system as a whole.

In close analogy to the quantum situation, where measurement separates ontic (holistic) and epistemic 
(local) levels of description, the appearance of  
conscious contents as manifestations of unconscious material must be considered as a  
transformation between fundamentally different mental modalities. The unconscious mode is left
(and maybe even changed) whenever a conscious content emerges out of it. Long ago, James (1950) 
perfectly paraphrased this situation by the impossibility to recognize what darkness is by switching 
the light on.

This difference between the two modalities becomes particularly interesting if the (unconscious) global 
state is in fact an entangled state. The entanglement can then refer to unconscious personal material 
or to the unconscious of collectives, resembling a specific realization of Jung's concept of the collective 
unconscious. In the case of individuals and their unconscious, the global system
would correspond to some undifferentiated personal realm of the unconscious without local, separate 
categories, while elements of consciousness, such as mental categories, are local and separate.  
Particular mental categories (including the ``I'' or ``self'' as one of the
most significant among them) are conceived to emerge by the transformation of unconscious material 
into consciously and empirically accessible categories. 

In order to discuss transference and countertransference phenomena, it is necessary to 
address more than one individual. This makes it mandatory to consider the (collective) unconscious 
of a group, such as described in the examples above.  
If by some sort of ``organizational closure'' individuals establish a tightly bound system 
-- a pair of lovers, a family, or another social group -- then novel conscious contents can emerge 
at some particular part of the system (e.g., in one individual)
as a result of a manifestation of unconscious material within the system as a whole.   
Again, it should be emphasized that unconscious material is not simply ``made conscious'' as
it is; the emergence of conscious manifestations of unconscious material must be understood 
as a transition between fundamentally different mental modalities.   

Conversely, if the binding is intense enough, personal unconscious material of one individual can 
become part of the collective unconscious of the system as a whole by some kind of composition
(rather than decomposition)  
process. In this way the collective unconscious of the system as a whole becomes a ``melting
pot'' of highly correlated individual contributions, to be formally described as an entangled state. 

Although the mechanisms of decomposition and composition for such a scenario are far from
being explored in detail, the basic framework of weak quantum theory offers an interesting perspective for
what can happen in transference and countertransference processes even beyond therapeutic
applications.  For instance, if in a marriage relationship one of 
the partners ``experiences'' something which is systemically unconscious,
say the wish to separate, then this wish can manifest itself in the other partner's awareness 
as his or her own wish. The fatal aspect of such a phenomenon is that the corresponding
material is mostly taken at face value rather than as a possible indicator of something 
originally belonging to another person. 

\section{Summary}

The core content of this paper is the formulation of a weak version of quantum
theory. It is motivated by the attempt to find a formal framework for addressing the 
concepts of complementarity and entanglement not only within the context of ordinary
quantum physics, but also in more
general contexts. The weak version of quantum theory is based on a minimal set of
axioms. The basic structure of the resulting mathematical framework is that of a
monoid. 
   
Ordinary quantum theory can be recovered from this framework by additional axioms,
restrictions, and specifications. For example, the weak version does not necessarily entail 
a Hilbert space 
representation or a probabilistic interpretation. The non-commutativity of observables
is not necessarily quantified by Planck's constant. Bell-type inequalities cannot necessarily be
formulated in weak quantum theory.

Among the many examples for complementary relations that can be found in the literature,
two case studies were presented to demonstrate the applicability of weak quantum
theory. They refer to (1) complementary types of dynamical descriptions of physical systems,
and (2) the 
relation between conscious and unconscious processes in psychoanalytic and psychotherapeutic
settings.

These examples show that there are different levels of generalization between weak quantum theory
and ordinary quantum theory, depending on which restrictions are added to the minimal,
weak framework. While example (2) is likely to need the full generalization 
of the weak version, 
only a few conditions of ordinary quantum theory are relaxed in example (1). 
These conditions are discussed to some detail. 

The main benefit of weak quantum theory is its applicability beyond physics. 
This is possible since the generalized formal framework is conceptually decoupled from
its application to ordinary quantum physics. Although ``naive'' analogies to the physical 
domain remain helpful, the wider scope allows a formally based approach to non-physical
situations as well. Clearly there may be situations which are far from a formal description
in terms of non-commutative operators as yet. On the other hand, there are concrete proposals,
e.g.~for cognitive processes (cf.~Gernert 2000) toward corresponding descriptions which 
underline the potential usefulness of  weak quantum theory.

Different future steps to explore this potential usefulness are conceivable.
One of them is a more formal and detailed discussion of applications, which would go beyond 
the frame of the present paper. One example concerning placebo research was mentioned
already. Another, much more general problem refers to the question how psychophysical
relationships could be treated within weak quantum theory. 
Needless to say, it will be crucial to propose and carry out experiments demonstrating the 
full power of the approach.  

\section*{Acknowledgments}

We are grateful to Klaus Jacobi for sharing his philosophical expertise in inspiring
discussions. We also appreciate helpful comments by three anonymous referees.

\section*{References}

\begin{enumerate}

\item A.~Aspect, J.~Dalibard, and G.~Roger (1982): Experimental test of Bell's
      inequalities using time-varying analyzers. {\it Phys.~Rev.~Lett.}
      {\bf 49}, 1804--1807 (1982).

\item H.~Atmanspacher (1991a): A propositional lattice for the 
      logic of temporal predictions. In {\it Solitons and Chaos}. 
      Edited by I.~Antoniou and F.J.~Lambert. Springer, Berlin, 1991, 
      pp.~58--70. 

\item H.~Atmanspacher (1991b): Incommensurability of Liouvillean dynamics 
      and information dynamics.  In {\it Parallelism, Learning, 
      Evolution}. Edited by 
      J.D.~Becker, I.~Eisele, and F.W.~M\"undemann. Springer, Berlin, 
      1991, pp.~482--499.

\item H.~Atmanspacher (1997): Dynamical entropy in dynamical systems.
      In {\it Time, Temporality, Now}. Edited by H.~Atmanspacher and E.~Ruhnau.
      Springer, Berlin, 1997, pp.~327--346. 

\item H.~Atmanspacher (2000): Ontic and epistemic descriptions of chaotic systems.
      In {\it Computing Anticipatory Systems}. Edited by D.~Dubois.
      American Institute of Physics, New York, 2000, pp.~465--478.

\item H.~Atmanspacher (2001): Mind and matter as asymptotically disjoint,
      inequivalent representations with broken time-reversal symmetry.  In press.

\item H.~Atmanspacher and H.~Primas (1996): The hidden side of Wolfgang Pauli.
      {\it J.~Consc.~Studies} {\bf 3}, 112--126 (1996).

\item H.~Atmanspacher and H.~Scheingraber (1987): A fundamental link between
      system theory and statistical mechanics. {\it Found.~Phys.} {\bf 17},
      939--963 (1987).   

\item J.S.~Bell (1964): On the Einstein Podolsky Rosen paradox. {\it Physics}
      {\bf 1}, 195--200 (1964).

\item P.~Bernays (1948): \"Uber die Ausdehnung des Begriffes der Komplementarit\"at
      auf die Philosophie. {\it Synthese} {\bf 7}, 66--70 (1948). 

\item G.~Birkhoff and J.~von Neumann (1936): The logic of quantum mechanics. 
       {\it Ann.~Math.} {\bf 37}, 823--843 (1936).

\item N.~Bohr (1948): On the notions of causality and complementarity.
     {\it Dialectica} {\bf 2}, 312--319 (1948).

\item D.J.~Chalmers (1996): {\it The Conscious Mind}. Oxford University Press,
      Oxford, 1996. 

\item A.~Einstein, B.~Podolsky, and N.~Rosen (1935): Can quantum-mechanical 
      description of physical reality be considered complete? 
      {\it Phys.~Rev.} {\bf 47}, 777--780 (1935).



\item S.~Freud (1992): {\it Vorlesungen zur Einf\"uhrung in die Psychoanalyse}.
         Fischer, Frankfurt 1992.

\item D.~Gernert (2000): Towards a closed description of observation processes.
         {\it BioSystems} {\bf 54}, 165--180 (2000). 

\item P.~Grassberger and I.~Procaccia (1983): Estimation of the Kolmogorov entropy 
      from a chaotic signal. 
      {\it Phys.~Rev.~A} {\bf 28}, 2591--2593 (1983). 

\item K.~Gustafson and B.~Misra (1976): Canonical commutation relations of 
      quantum mechanics and stochastic regularity. {\it Lett.~Math.~Phys.} 
      {\bf 1}, 275--280 (1976). 

\item R.~Haag (1996): {\it Local Quantum Physics}. Springer, Berlin, 1996.

\item W.~James (1950): {\it The Principles of Psychology, Vol.~1}. Dover,
      New York, 1950, Chap.~IX.

\item M.~Jammer (1974): {\it The Philosophy of Quantum Mechanics}. Wiley,
      New York, 1974, Chaps.~5 and 6.

\item P.~Jordan (1947): {\it Komplementarit\"at und Verdr\"angung}. Strom
      Verlag, Hamburg, 1947.

\item C.G.~Jung (1921): {\it Psychologische Typen}. Rascher, Z\"urich, 1921. 

\item C.G.~Jung (1971): Theoretische \"Uberlegungen zum Wesen des
       Psychischen. In {\it Gesammelte Werke, Band 8.} Walter, Olten, footnote
       129, pp.~261f. English translation: 
       On the nature of the psyche. In {\it Collected
       Works, Vol.~8}. Princeton University Press, Princeton, 1969,  
       footnote 130, pp.~229f.  

\item C.G.~Jung and W.~Pauli (1952): {\it Naturerkl\"arung und Psyche}.
      Rascher, Z\"urich, 1952.  



\item K.~Kornwachs and W.~von Lucadou (1985): Pragmatic information as a 
        nonclassical concept to describe cognitive systems. {\it Cogntive Systems}
        {\bf 1}, 79--94 (1985). 

\item F.R.~Krueger (1984): {\it Physik und Evolution}. Parey, Berlin, 1984. 

\item P. Kruse, M. Stadler (1995): {\it Ambiguities in Mind and Nature}, Springer, Berlin, 1995.


\item K.M.~Meyer-Abich (1965): {\it Korrespondenz, Individualit\"at und Komplementarit\"at}.
      Steiner, Wiesbaden, 1965.

\item B.~Misra (1978): Nonequilibrium entropy, Lyapounov variables, and
      ergodic properties of classical systems.
      {\it Proc.~Ntl.~Acad.~Sci.~USA} {\bf 75}, 1627--1631 (1978).

\item B.~Misra, I.~Prigogine, and M.~Courbage (1979): From deterministic dynamics 
     to probabilistic descriptions. {\it Physica A} {\bf 98}, 1--26 (1979); 
      cf.~A.S.~Wightman; {\it Math.~Rev.} {\bf 82e}, 58066 (1982). 
 
\item B.~Misra and I.~Prigogine (1983): Irreversibility and nonlocality. 
      {\it Lett.~Math.~Phys.} {\bf 7}, 421--429 (1983).

\item D.~Murdoch (1987): {\it Niels Bohr's Philosophy of Physics}.
      Cambridge University Press, Cambridge, 1987.

\item A.~Pais (1991): {\it Niels Bohr's Times In Physics, Philosophy, and Politics}.
      Clarendon, Oxford, 1991. 

\item W.~Pauli, ed.~(1948): Special issue on Complementarity.
     {\it Dialectica} {\bf 2} (1948).

\item W.~Pauli (1954): Naturwissenschaftliche und erkenntnistheoretische Aspekte
      der Ideen vom Unbewussten. {\it Dialectica} {\bf 8}, 283--301 (1954).

\item C.~Piron (1976): {\it Foundations of Quantum Physics}. Benjamin,
        Reading, 1976. 

\item E.~Plaum (1992): Bohrs quantentheoretische Naturbeschreibung und die
      Psychologie. {\it Psychologie und Geschichte} {\bf 3}, 94--101 (1992).

\item H.~Primas (1983): {\it Chemistry, Quantum Mechanics, and Reductionism}.
      Sprin\-ger, Berlin, 1983.

\item H.~Primas (1990): Mathematical and philosophical questions in the theory of
      open and macroscopic quantum systems. In {\it Sixty-Two Years of
      Uncertainty}, ed.~by A.I.~Miller, Plenum, New York, 1990,
      pp.~233--257.


\item E.~Scheibe (1973): {\it The Logical Analysis of Quantum Mechanics}.
      Pergamon, Oxford, 1973, pp.~82--88.

\item E.~Schr\"odinger (1935): Die gegenw\"artige Situation in der
      Quantenmechanik. {\it Naturwiss.~23}, 807--812, 823--828, 844--849. 

\item R.~Shaw (1981): Strange attractors, chaotic behavior, and information flow. 
      {\it Z.~Naturforsch.} {\bf 36a}, 80--112 (1981). 

\item W.~Thirring (1981): {\it Quantum Mechanics of Atoms and Molecules}. Springer, 
         Berlin, 1981.

\item D.~Tj\o stheim (1976): A commutation relation for wide sense stationary
      processes. {\it SIAM J.~Appl.~Math.} {\bf 30}, 115--122 (1976).

\item M.~Varga von Kibed (1998): Bemerkungen \"uber philosophische Grundlagen
         und methodische Voraussetzungen zur systemischen Aufstellungsarbeit.
          In {\it Praxis des Familien-Stellens}, ed.~G.~Weber, Carl Auer, Heidelbert, 1998.
         Vergleiche weitere Beitr\"age des Autores im gleichen Band.


\item H.~Walach and H.~R\"omer (2000): Complementarity is a 
          useful concept for consciousness studies. A reminder. {\it Neuroendocrinology 
          Letters 21}, 221--232.


\item E.~von Weizs\"acker (1974): Erstmaligkeit und Best\"atigung als
       Komponenten der pragmatischen Information, in {\it Offene
       Systeme I}, ed.~by E.~von Weizs\"acker. Klett-Cotta, Stuttgart, 1974,
       pp.~83--113.

\end{enumerate}

\end{document}